\documentclass[12pt]{spieman}  
\usepackage{amsmath,amsfonts,amssymb}
\usepackage{graphicx}
\usepackage{setspace}
\usepackage{tocloft}
\usepackage{lineno}
\usepackage{array}
\newcommand{\PreserveBackslash}[1]{\let\temp=\\#1\let\\=\temp}
\newcolumntype{C}[1]{>{\PreserveBackslash\centering}p{#1}}
\newcolumntype{R}[1]{>{\PreserveBackslash\raggedleft}p{#1}}
\newcolumntype{L}[1]{>{\PreserveBackslash\raggedright}p{#1}}
\usepackage{caption} 
\usepackage{ulem}
\usepackage{xcolor}

\newcommand{\change}[1]{\textcolor{black}{#1}}
\newcommand{\electron}{$\text{e}^-$}

\title{Modeling and performance analysis of Implicit Electric Field Conjugation with two deformable mirrors applied to the Roman Coronagraph}

\author[a]{Kian Milani}
\author[b]{Ewan S. Douglas}
\author[b]{Sebastiaan Y. Haffert}
\author[b]{Kyle Van Gorkom}
\affil[a]{University of Arizona, Wyant College of Optical Sciences, Tucson, Arizona, United States}
\affil[b]{University of Arizona, Steward Observatory, Tucson, Arizona, United States}

\cftpagenumbersoff{figure}
\cftpagenumbersoff{table} 
\begin{document} 
\maketitle

\begin{abstract}
High-order wavefront sensing and control (HOWFSC) is key to create a dark hole region within the coronagraphic image plane where high contrasts are achieved. The Roman Coronagraph is expected to perform its HOWFSC with a ground-in-the-loop scheme due to the computational complexity of the Electric Field Conjugation (EFC) algorithm. This scheme provides the flexibility to alter the HOWFSC algorithm for given science objectives. The baseline HOWFSC scheme involves running EFC while observing a bright star such as $\zeta$ Puppis to create the initial dark hole followed by a slew to the science target.

The new implicit EFC (iEFC) algorithm removes the optical diffraction model from the controller, making the final contrast independent of model accuracy. While previously demonstrated with a single DM, iEFC is extended to two deformable mirror systems in order to create annular dark holes. First, an overview of both EFC and iEFC is presented. The algorithm is then applied to the Wide-Field-of-View Shaped Pupil Coronagraph (SPC-WFOV) mode designed for the Roman Space Telescope using end-to-end physical optics models. Initial noiseless monochromatic simulations demonstrate the efficacy of iEFC as well as the optimal choice of modes for the SPC-WFOV instrument. Further simulations with a 3.6\% wavefront control bandpass and a broader 10\% bandpass then demonstrate that iEFC can be used in broadband scenarios to achieve contrasts below $10^{-8}$ with Roman. Finally, \change{an EMCCD model is implemented to estimate calibration times and predict the controller's performance. Here,} $10^{-8}$ \change{contrasts are achieved} with a calibration time of about 6.8 hours assuming the reference star is $\zeta$ Puppis. The results here indicate that iEFC can be a valid HOWFSC method that can mitigate the risk of model errors associated with space-borne coronagraphs, but to maximize iEFC performance, lengthy calibration times will be required to mitigate the noise accumulated during calibration.
\end{abstract}


{\noindent \footnotesize\textbf{*}Kian Milani,  \linkable{kianmilani@arizona.edu} }

\begin{spacing}{1}   

\section{Introduction}
\label{sec:intro}
With thousands of exoplanets having been discovered primarily via indirect detection methods including transit photometry and radial velocity measurements, further studies are desired to understand the dynamics of planet formation, interactions with debris disks, and aid the discovery of potentially habitable worlds. These studies can be enabled if the challenges of direct-imaging are overcome, of which the two primary challenges are the relatively small angular separation between the host star and the exoplanet as well as the small planet to star flux ratio. For perspective, a Sun-Earth analog at 10 parsecs will have an angular separation of about 0.1 arcsec and a flux ratio of about $10^{-10}$\cite{woolf-earth-like-contrast-1998}. While coronagraphs and high-order wavefront control algorithms have been demonstrated under laboratory conditions with ground-based \change{instruments\cite{potier-vlt-sphere-efc, ahn-combining-efc-ldfc}}, atmospheric turbulence, telescope stability, and absorption in certain bandpasses limit ground-based performance. A key milestone for direct-imaging will be the launch of the Nancy Grace Roman Space Telescope, which will carry an onboard coronagraph with multiple modes for imaging and spectral characterization\cite{noecker-wfirst-afta-2016}. 

While acting as a technology demonstration, the Roman Coronagraph is expected to bridge the performance gap between current high-contrast imaging capabilities and what is expected for a future 6 meter class Habitable Worlds Observatory (HWO) recommended by the decadal survey to image Earth-like exoplanets\cite{astro2020decadal}. Predicted to achieve contrasts on the order of $10^{-9}$,\cite{Bailey-roman-cgi-2023} the Coronagraph Instrument is particularly focused on demonstrating two distinct coronagraph designs, low-noise electron multiplying detectors, and high-order wavefront sensing and control (HOWFSC) algorithms required to suppress speckles from quasi-static aberrations within the optical train. The primary method intended for the HOWFSC scheme has been the use of model-based Electric Field Conjugation (EFC)\cite{give'on-bb-wavefront-correction}. Described in Section \ref{sec:efc}, EFC is combined with Pairwise Probing (PWP) in order to sense and minimize the focal plane electric field using an instrument model. 

Currently, EFC is expected to achieve the desired contrasts for Roman. However, the contrast attained by EFC has been demonstrated to degrade, in some cases by an order of magnitude, due to model errors, particularly those associated with the DM translation and rotation\cite{potier-comparing-fpwfs}. This is the motivation to study the alternative implicit EFC method \cite{haffert-iefc} where the Jacobian is constructed entirely from empirical data and inherently includes all perturbations in the optical system. For a space-borne telescope, iEFC can be a particularly valuable method as it can mitigate the risk associated with instrument performance in the presence of instrument uncertainties. In addition, iEFC does not require the focal plane electric field to be sensed prior to computing the actuator commands. This simplifies the controller because only a single regularization parameter needs to be optimized whereas an implementation of EFC with PWP requires two regularization parameters, one for the initial electric field estimation during PWP and one for the conjugation using EFC\cite{haffert-iefc}.

Here, iEFC is extended to use 2 DMs and proven capable of creating an annular dark hole such that it can be applied to the Roman Coronagraph as well as designs for future missions such the HWO. The Wide-Field-of-View Shaped Pupil Coronagraph (SPC-WFOV) mode is chosen because preliminary simulations demonstrated that iEFC is likely not suited to the Hybrid Lyot Coronagraph (HLC) due to the frequent relinearization of the Jacobian required for the HLC if the DM design patterns are not used\cite{milani-cgi-iefc-2023} or are inaccurate\cite{zhou-cgi-howfsc-2020}. Specifically, the empirical calibration of the iEFC Jacobian would likely make repeated relinearizations unfeasible for the HLC because of the required duration of each Jacobian calibration. 

Section \ref{sec:defs} provides an overview of the EFC and iEFC algorithms, including how an arbitrary basis of modes can be used for each method. Additionally, iEFC is extended to use two DMs and initially demonstrated with a scalar vortex coronagraph (SVC) model. Section \ref{sec:motive} then includes a demonstration of EFC for the SPC-WFOV mode that illustrates how instrument perturbations unaccounted for in a model result in contrast degradation and a more complex control scheme. \change{In Section \ref{sec:mono-iefc}, the efficacy of iEFC is tested with monochromatic simulations} using multiple modal bases to demonstrate which is optimal for the SPC-WFOV instrument. Section \ref{sec:broad-iefc} then expands the iEFC simulations for a single narrowband wavefront control bandpass and a broader 10\% bandpass. Lastly, Section \ref{sec:noisy-iefc} integrates flux estimates for a reference star and an EMCCD model into the simulations to test the feasibility of iEFC for both bandpasses considered. The results demonstrate that iEFC can be a suitable method for the SPC-WFOV mode and can mitigate the risk of pupil misalignments after the launch of Roman. However, the disadvantage of iEFC will be noise accumulated during calibration and the integration time required to acquire all calibration frames. Given EFC has been rigorously tested and simulated for the Roman Coronagraph and is expected to achieve the desired contrast goals, iEFC should likely be tested during the latter stages of the Roman mission in order to further study the efficacy of the method. This would inform future missions, ideally with less obstructed and larger apertures capable of capturing more photons to reduce calibration times and mitigate the effects of calibration noise.

\section{Definitions and Algorithm}
\label{sec:defs}
For clarity between the EFC and iEFC, Table \ref{tab:vars} in Appendix \ref{app:defs} has a list of variables used for the formalism of both methods. Variables in bold font represent column vectors, scalars are all lowercase variables, and \change{matrices and scalar fields} are all uppercase variables. The precise shape of all matrices is also included Table \ref{tab:vars}.

The images throughout these simulations are presented in units of normalized intensity (NI), defined as 
\begin{equation}
    \change{I_N(x,y)} = \frac{I_{image}(x,y)}{max(I_{uo}(x,y))}
\end{equation}
\noindent where $I_{image}(x,y)$ is the intensity of a raw coronagraphic image and $I_{uo}(x,y)$ is the intensity of \change{the unocculted on-axis source (no focal plane mask in the optical train).} While not a direct measurement of contrast, this metric is closely related and is the same used in FALCO\cite{sidick-falco-2}. When contrast is referenced in latter sections, it will refer to this definition of normalized intensity.

\subsection{EFC}
\label{sec:efc}
Following the formalism of EFC from Give'on et al.\cite{give'on-bb-wavefront-correction}, the fundamental concept is that the irradiance in the image plane can be minimized by computing DM commands that destructively interfere with the measured electric field\cite{give'on-bb-wavefront-correction}. This is done by using a model of the instrument to estimate the electric field response of DM modes \change{in} the image plane. For a coronagraph, this electric field may be written as some linear operation $C[\text{\space}]$ acting on the pupil plane wavefront $A(x,y)e^{i\Phi(x,y)}$ where $A(x,y)$ and $\Phi (x,y)$ are the \change{scalar fields} corresponding to the pupil plane amplitude and phase aberrations respectively. 

\begin{equation}
    \label{eq:efc-1}
    \change{E_{im}(x,y)} = C[A(x,y)e^{i\Phi(x,y)}]
\end{equation}

With coronagraphs comprising multiple relays between pupil planes and focal planes, $C[\text{\space}]$ is often a series of Fourier Transforms propagating the wavefront between the transverse planes. For a system with a single DM, the electric field can be written as 

\begin{equation}
    \label{eq:efc-2}
    \change{E_{im}(x,y)} = C[A(x,y)e^{i\Phi(x,y)}e^{i\Phi_{DM}(x,y)}]\text{.}
\end{equation}

\noindent Here, $\Phi_{DM}(x,y)$ is the phase induced by the surface of the DM expressed as a sum of weighted influence functions $F_n(x,y)$ with the expression 

\begin{equation}
    \Phi_{DM}(x,y) = \frac{4\pi}{\lambda}\sum_{n=1}^{N_{acts}} a_n F_n(x,y)\text{.} 
\end{equation} 

\noindent Assuming that the DM surface imparts a small phase, the DM phasor can be approximated as $e^{i\Phi_{DM}(x,y)} \approx 1 + \change{i\Phi_{DM}(x,y)}$, allowing for the following simplification to the image plane electric field.

\begin{equation}
    \label{eq:efc-3}
    \change{E_{im}(x,y)} = C[A(x,y)e^{i\Phi(x,y)}] + iC[\Phi_{DM}(x,y)]
\end{equation}

Once sampled by the science camera detector, \change{$E_{im}(x,y)$} is written as the concatenated vector $\mathbf{E_{im}}$ containing the real and imaginary components of the electric field at each pixel in the dark hole. The individual terms for the electric field contributed by the system aberrations and the DM influence functions are now written as 
\begin{equation}
    \change{C[A(x,y)e^{i\Phi(x,y)}] = \mathbf{E_{ab}}}
\end{equation}
\noindent and 
\begin{equation}
    \change{iC[\Phi_{DM}(x,y)] = G_1\mathbf{A_1}}
\end{equation}
\noindent respectively. Here, $G_1$ is the Jacobian that transforms the actuator heights of DM1 ($\mathbf{A_1}$) into the electric field contributed by the DM at the focal plane. With this, the image plane electric field is written as the following system of linear equations. 

\begin{equation}
    \label{eq:efc-3}
    \mathbf{E_{im}} = \mathbf{E_{ab}} + G_{1} \mathbf{A_1}
\end{equation}

As presented in \change{Give'on et al\cite{give'on-bb-wavefront-correction}}, when a second DM is used outside a pupil plane, this system of linear equations is expanded to include the term $G_2 \mathbf{A_2}$. While similar to the Jacobian for the pupil plane DM, the Jacobian $G_2$ includes the angular spectrum propagation affects from the plane of the second DM back to the pupil plane. As a system of linear equations, the two Jacobians can be concatenated to capture the influence of all DM actuators such that the final equation for the image plane electric field is simplified to

\begin{equation}
    \label{eq:efc-3}
    \begin{split}
    \mathbf{E_{im}} = \mathbf{E_{ab}} + G_1\mathbf{A_1} + G_2 \mathbf{A_2} = \mathbf{E_{ab}} + G_{EFC} \mathbf{A} 
    \\
    \text{where } G_{EFC} = [G_1, G_2] \text{ and } \mathbf{A} = \begin{bmatrix}\mathbf{A_1} \\\mathbf{A_2}\end{bmatrix}
    \end{split}
\end{equation}

In the case where an arbitrary basis of DM modes are used instead of individual actuators, this can be written more generally in terms of the modal coefficients $\mathbf{m_{c}}$. The basis of DM modes chosen are referred to as the calibration modes that are contained in the matrix $M_{modes}$. Note that the shape of the Jacobian will depend on the amount of calibration modes used and the equation for the image plane electric field is then

\begin{equation}
    \mathbf{E_{im}} = \mathbf{E_{ab}} + G_{EFC} \mathbf{m_c} \text{.}
\end{equation}

\begin{figure}[H]
    \centering
    \includegraphics[scale=0.73]{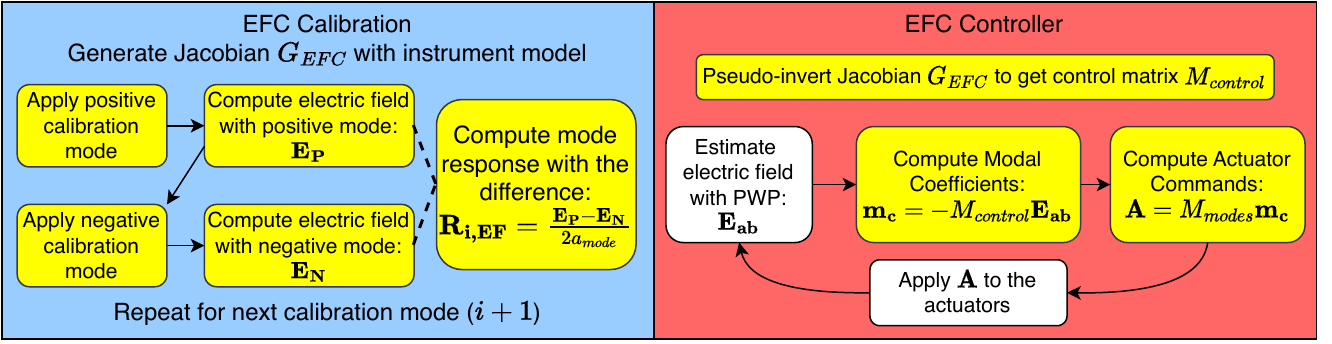}
    \caption{This diagram illustrates the functional steps for model-based EFC. The Jacobian $G_{EFC}$ is constructed by applying positive and negative calibration modes within the model and calculating the electric field at the focal plane. The response/derivative is computed from the difference normalized by the amplitude of the mode applied ($a_{mode}$). The EFC controller uses the pseudo-inverted Jacobian to compute the update to the DM actuators using the estimated electric field $\mathbf{E_{ab}}$. All boxes in yellow indicate a step that is performed computationally.}
    \label{fig:efc-flowchart}
\end{figure}

Figure \ref{fig:efc-flowchart} illustrates the typical practices for how EFC is calibrated and performed. Prior to the EFC control loop, the Jacobian is computed using the central difference approximation \change{in Equation \ref{eq:efc-response} to estimate the response of the electric field in the image plane for each calibration mode. To do so,} the positive and negative calibration modes \change{are applied in a Fourier based optical model with $a_{mode}$ being the chosen calibration amplitude to calculate the fields referenced as} $\mathbf{E_P}$ and $\mathbf{E_N}$ respectively. \change{While it is possible to estimate the response using the nominal electric field and only applying a single positive or negative calibration mode, the central difference approximation decreases the error from nonlinear terms\cite{will-adefc-2021}. Each response stored in the Jacobian as a single vector $\mathbf{R_{i,EF}}$ where $i$ denotes the index of the calibration mode.} 

\begin{equation}
    \label{eq:efc-response}
    \change{\mathbf{R_{i,EF}} = \frac{\mathbf{E_P}-\mathbf{E_N}}{2a_{mode}}}
\end{equation}

\begin{equation}
    G_{EFC} = 
    \begin{bmatrix}
        \mathbf{R_{1,EF}} & \mathbf{R_{2,EF}} & ... & \mathbf{R_{i,EF}} & ... & \mathbf{R_{N_{modes}, EF}}
    \end{bmatrix}
\end{equation}

During each iteration of EFC, the solution for modal coefficients to minimize the electric field are commonly found by minimizing a cost function such as 

\begin{equation}
    J = |\mathbf{E_{ab}} + G_{EFC}\mathbf{m_c}|^2  + \lambda|\mathbf{m_c}|^2
\end{equation}

\noindent where $\lambda$ is a regularization parameter penalizing the use of large actuator strokes. Throughout the simulations presented, the ``$\beta$ regularization'' technique described in Sidick et al
\cite{sidick_optimizing_2017} is used for both EFC and iEFC simulations. With this technique, the regularization parameter chosen by the user is $\beta$. Once specified, the control matrix ($M_{control}$) is computed numerically using the equation 
\begin{equation}
    M_{control} = (G_{EFC}^\top G_{EFC} + \alpha^2 10^\beta I)^{-1} G_{EFC}^\top  \text{.}
\end{equation}
\noindent Here, $\alpha$ is the maximum singular value of the square matrix $G_{EFC}^\intercal G_{EFC}$. The modal coefficients that minimize the cost function are calculated with the matrix-vector product $-M_{control}\mathbf{E_{ab}}$. These modal coefficients can be transformed into actuator heights using $M_{modes}$ as demonstrated in the following equation. 

\begin{equation}
    \mathbf{A} = 
    \begin{bmatrix}
        \mathbf{A_{1}} \\
        \mathbf{A_{2}}
    \end{bmatrix} = M_{modes} \mathbf{m_c} =  - M_{modes} M_{control} \mathbf{E_{ab}}
\end{equation}

A critical component to EFC is how the electric field in the focal plane is estimated prior to the actuator command being calculated. While the Self Coherent Camera (SCC) is a method of estimating the focal plane electric field using interference from a reference channel\cite{baudoz2005self}, the common technique proposed for Roman is Pairwise Probing (PWP)\cite{krists-bible}. This method uses at least two linearly independent DM probe commands to generate phase diversity within the focal plane. For each probe command, two images are recorded, one with the positive probe and one with the negative probe. The difference images of each probe command are then used to estimate the electric field in each pixel of the desired dark hole by solving \change{Equation \ref{eq:pwp} for $\mathbf{E_{ab}}$ where $\boldsymbol{\delta}$ is the concatenated vector of difference images and $M_{probes}$ is a matrix with the real and imaginary components of the electric field contributed by the DM probes estimated using the instrument model. Solving this system of linear equations requires another pseudo-inverse in which the regularization parameter $\lambda_{PWP}$ is used to make the algorithm robust against noise.} 

\begin{equation}
\label{eq:pwp}  
\change{\boldsymbol{\delta}} = M_{probes} \mathbf{E_{ab}} 
\end{equation} 

\begin{equation}
\label{eq:pwp-solution}  
\change{\mathbf{E_{ab}} = (M_{probes}^\top M_{probes} + \lambda_{PWP} I )^{-1} M_{probes}^\top \boldsymbol{\delta}}
\end{equation} 

Various other implementations of PWP have been developed, including extensions to broadband estimation\cite{redmond-pwp} and implementations of a Kalman filter for improved estimation\cite{groff-pwp-kf}, but all rely on the difference images from pairwise probes. 

\subsection{iEFC}
\label{sec:iefc}
As described in Haffert et al\cite{haffert-iefc}, iEFC includes no explicit calculation of the electric field within each iteration because the electric field is not the the desired control variable, rather, difference images of probes are the control variable. The formalism of PWP demonstrates that difference images of two or more linearly independent probes are related to the focal plane electric field by a linear transformation making them a linear proxy for the electric field. Therefore, minimizing the difference images also minimizes the electric field. 

Because the control variable for iEFC is measurable, a Fourier optics model of the instrument is not required to compute the Jacobian. Instead, iEFC is calibrated by measuring the response $\mathbf{R_{i,\delta}}$ for each calibration mode. Here, the response measured for a single mode is another central difference of the measurements for both the positive and negative calibration mode. Explained by Equation \ref{eq:diffs}, the individual measurements themselves are the difference images $\boldsymbol{\delta}_{+j}$ and $\boldsymbol{\delta}_{-j}$. The value of $j$ references which probe the images are acquired for, $+$ and $-$ reference the sign of the calibration mode, while $P$ and $N$ reference the sign of the probe command. Each measured response is now a double difference that is stored into the iEFC Jacobian $G_{IEFC}$.

\begin{equation}
    \label{eq:diffs}
    \boldsymbol{\delta}_{+j} = \frac{\mathbf{I}_{+jP} - \mathbf{I}_{+jN}} { 2 a_{probe} } \text{ and }
    \boldsymbol{\delta}_{-j} = \frac{\mathbf{I}_{-jP} - \mathbf{I}_{-jN}} { 2 a_{probe} }
\end{equation}

\begin{equation}
    \label{eq:diffs}
    \boldsymbol{\delta}_{+} = 
    \begin{bmatrix}
        \boldsymbol{\delta}_{+1} \\
        \boldsymbol{\delta}_{+2} 
    \end{bmatrix} 
    \text{ and }
    \boldsymbol{\delta}_{-} = 
    \begin{bmatrix}
        \boldsymbol{\delta}_{-1} \\
        \boldsymbol{\delta}_{-2} 
    \end{bmatrix} 
\end{equation}

\begin{equation}
    \mathbf{R_{i,\delta}} = \frac{\boldsymbol{\delta}_{+}-\boldsymbol{\delta}_{-}}{2a_{mode}}
\end{equation}

\begin{equation}
    G_{IEFC} = 
    \begin{bmatrix}
        \mathbf{R_{1,\delta}} & \mathbf{R_{2,\delta}} & ... & \mathbf{R_{i,\delta}} & ... & \mathbf{R_{N_{modes},\delta}}
    \end{bmatrix}
\end{equation}

\begin{figure}[h]
    \centering
    \includegraphics[scale=0.73]{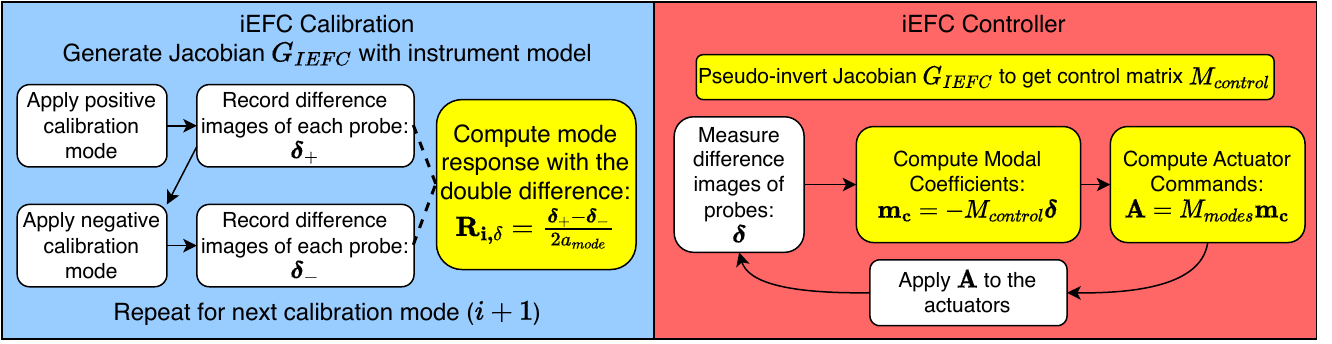}
    \caption{This diagram illustrates the critical steps for generating a Jacobian for iEFC by calibrating a chosen set of DM modes using difference images of probes concatenated into the vectors $\boldsymbol{\delta}_+$ and $\boldsymbol{\delta}_-$. By inverting the Jacobian and measuring $\boldsymbol{\delta}$ at each iteration, the difference images are minimized to reduce the electric field amplitude. Note that each measurement of $\boldsymbol{\delta}$ is also normalized by the amplitude of the probe ($a_{probe}$).}
    \label{fig:iefc-flowchart}
\end{figure}

Figure \ref{fig:iefc-flowchart} illustrates the procedures for calibrating iEFC and closing the loop. Similar to EFC, the cost-function can be defined such that the difference images will be minimized. This cost function can be used to then compute a control matrix for $M_{control}$. Each iteration of iEFC then consists of measuring the difference images $\boldsymbol{\delta}$ and solving for the modal coefficients by multiplying by $M_{control}$. 

\begin{equation}
    \label{eq:iefc-1}
    J = |\boldsymbol{\delta} + G_{IEFC} \mathbf{m_c}|^2  + \lambda|\mathbf{m_c}|^2
\end{equation}

\begin{figure}[h]
    \centering
    \includegraphics[scale=0.5]{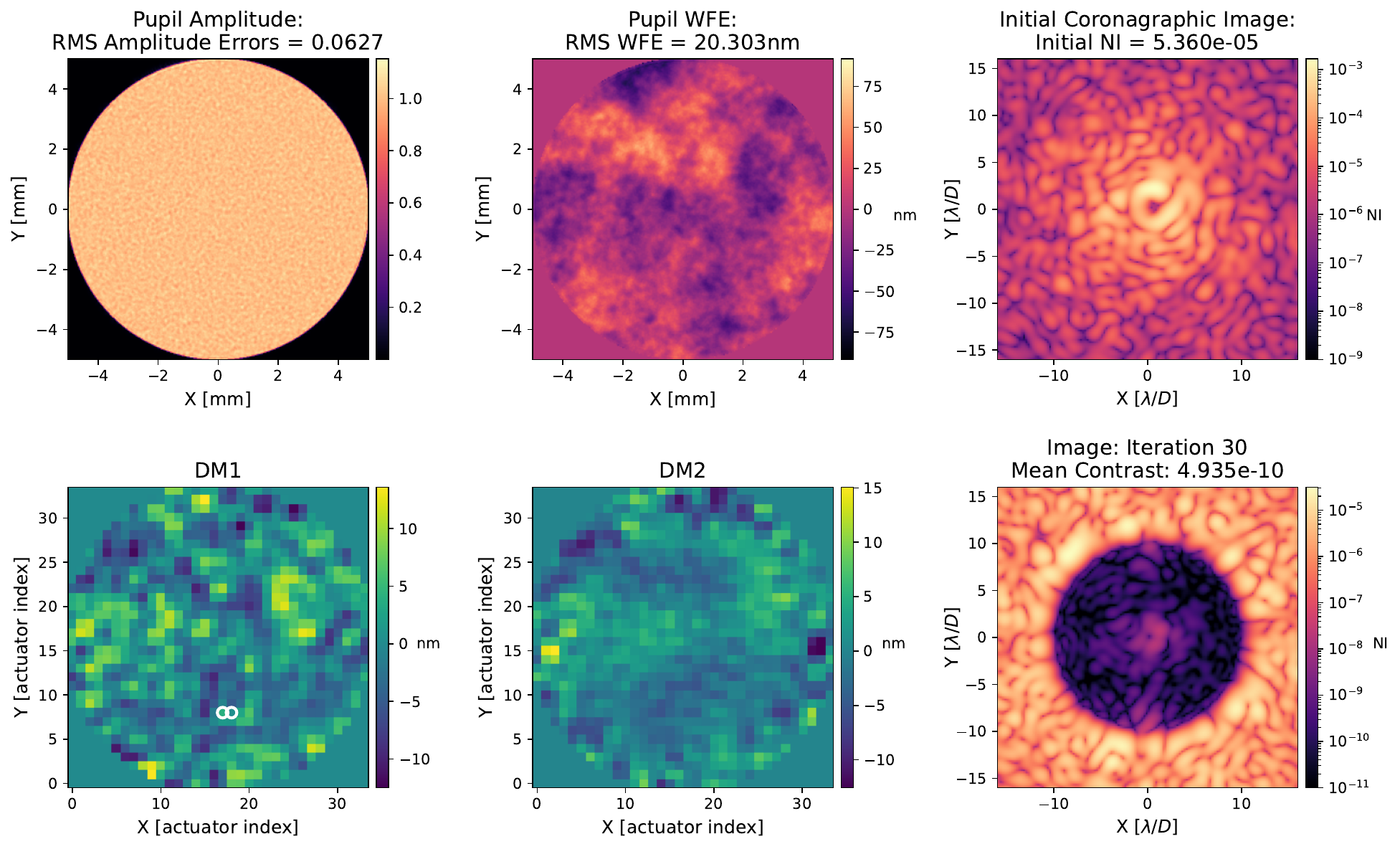}
    \caption{Annular dark hole created using iEFC with a scalar vortex coronagraph model where the final mean contrast is \change{$4.94\times10^{-10}$} after 30 iterations. The two white circles on the first DM indicate the actuators that are used as probes. }
    \label{fig:iefc-2dm-annular-dark-hole}
\end{figure}

The two DM iEFC controller is initially tested using an HCIPy\cite{por-hcipy-2018} scalar vortex coronagraph (SVC) model to verify the capability of creating an annular dark hole. The model uses a charge 6 vortex with a 90\% diameter Lyot stop. The wavelength for the simulation is 650nm with two $34\times34$ actuator DMs. The DM pupil is 10mm in diameter with 300mm separation between the DMs (roughly a Fresnel number of 512). Here, the control region spans from 2$\lambda/D$ to $10\lambda/D$. The calibration modes chosen are a basis of Fourier modes that span frequencies from 1.5$\lambda/D$ to $12\lambda/D$. The probes used are two single actuator pokes of adjacent actuators that are indicated by the white circles in Figure \ref{fig:iefc-2dm-annular-dark-hole} on DM1 (bottom left). The choice of Fourier modes, sampling, and probes will be expanded upon in Section \ref{sec:mono-iefc}. While this result is for an ideal monochromatic vortex coronagraph, the application of iEFC with 2 DMs is found to be suitable for annular dark holes.

\section{Roman SPC Physical Optics Model}
\label{sec:roman-pom}
The HOWFSC simulations performed here make use of an end-to-end physical optics model for the Roman Coronagraph. The model was created using POPPY\cite{perrin-poppy-2017} as the backend physical optics propagation software, but the prescription and data are the same as those in the roman\_phasec\_proper (v1.4) package\footnote{\href{https://sourceforge.net/projects/cgisim/}{https://sourceforge.net/projects/cgisim/}}. The motivation for choosing POPPY was to leverage end-to-end simulations entirely on a GPU as CuPy\cite{cupy_learningsys2017} was recently added as a computation feature in POPPY. The GPU computations greatly reducesimulation times and enable a more extensive investigation given the large number of computations required to simulate an iEFC Jacobian. When compared with results of the PROPER model, the mean normalized intensity of the POPPY model was found to be within a percent for an unocculted PSF and within a few percent for a coronagraphic image. More comparisons of the speckles and image morphology has been done in previous works\cite{milani-updated-roman-models}. 

\begin{figure}[h]
    \centering
    \includegraphics[scale=0.525]{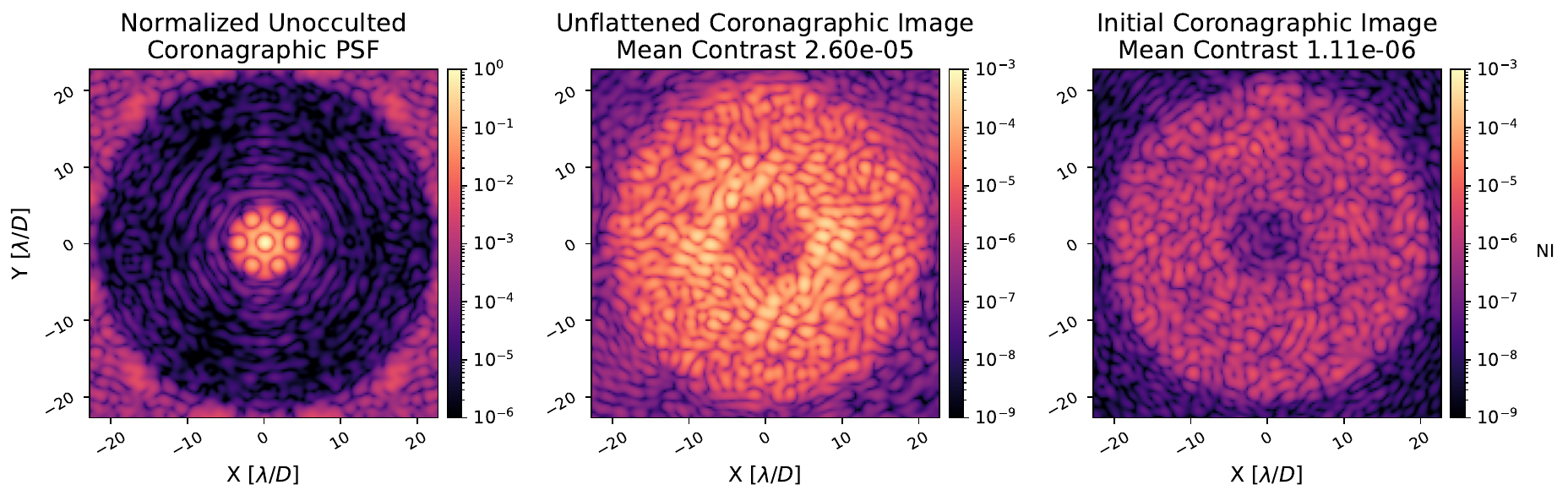}
    \caption{These images depict the initial state of the physical optics model prior to any HOWFSC. \change{The image of the unocculted on-axis source (left) is the reference image to which coronagraph frames are normalized. In the middle is the initial coronagraphic image without correction by a DM1 flatmap followed by the image on the right using the DM1 flatmap. This flatmap is used at the start of all simulations such that HOWFSC requires less stroke and attains better contrasts.}}
    \label{fig:spc-wide}
\end{figure}

Within the physical optics model, each optic includes a surface error map that is applied to the wavefront during propagation. In all EFC and iEFC simulations, the the initial state of the Fresnel model includes a DM1 flatmap correcting the pupil aberrations that improves the initial contrast for HOWFSC by more than an order of magnitude. Figure \ref{fig:spc-wide} demonstrates the initial images from the model prior to any HOWFSC. 

\section{Motivation for iEFC}
\label{sec:motive}
A risk for the Roman Coronagraph will be how well can the model be calibrated to the instrument once the instrument is in orbit. As such, many efforts have been made to investigate the impact of model errors. Previous simulations and experiments by Sidick et al\cite{sidick_optimizing_2017} and Marx et al\cite{marx_electric_2017} have demonstrated that using a $\beta$ scheduling approach can mitigate the impact of model errors because periodically relaxing the regularization parameter $\beta$ can improve the correction of poorly controlled modes even though overall contrast can degrade. By making the $\beta$ value more strict again, a better overall contrast can be attained. Effectively, this means model errors can be mitigated given a larger number of iterations to perform HOWFSC. 

\begin{figure}[h]
    \centering
    \includegraphics[scale=0.55]{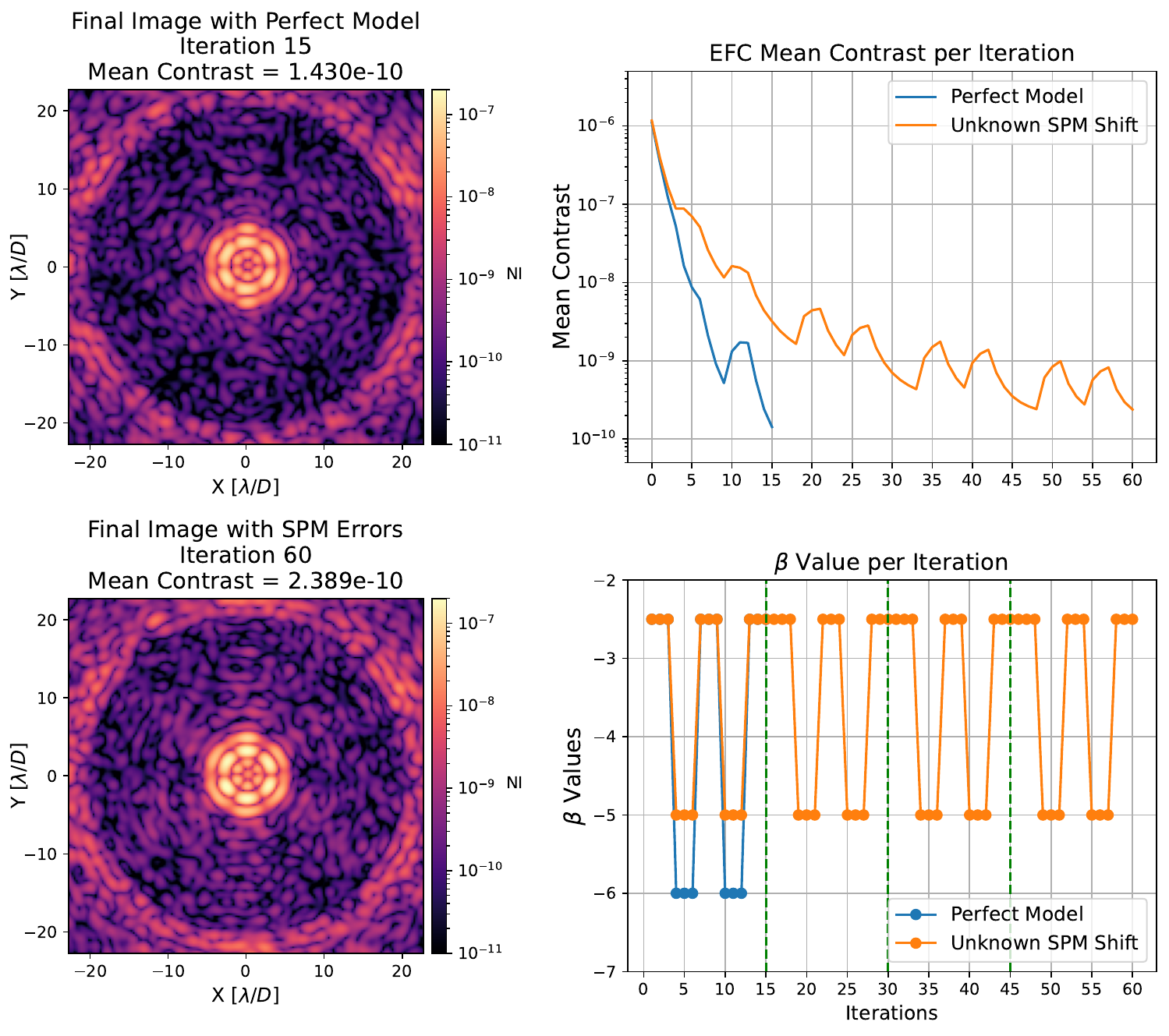}
    \caption{When no model errors are simulated, EFC converged to the result in the top-left after just 15 iterations. When introducing the SPM shear of about 85um, the result in the bottom-left was attained after 60 iterations. The top-right plot illustrates the mean contrast per iteration for both simulated scenarios. The bottom right plot presents the $\beta$ regularization value used during each iteration and the green lines represent each iteration at which the Jacobian was recomputed using the nonlinear coronagraph model.}
    \label{fig:efc-images}
\end{figure}

\change{Figure \ref{fig:efc-images} presents the results of a single case study is investigating the performance of the SPC-WFOV mode in the presence of an unknown lateral shear of the Shaped Pupil Mask (SPM)}. First, EFC is performed on the SPC-WFOV mode with no model errors. In this scenario, the end-to-end model is used to compute the Jacobian and the same model is then used within the EFC controller. Note that PWP is ignored by using the model to directly compute the electric field from which the \change{actuator commands are calculated}, so perfect electric field estimation is simulated. With no model errors, EFC rapidly converges to $1.43\times10^{-10}$ after just 15 iterations. 

For the scenario with an unknown lateral shear of the SPM, the model with no errors is again used to compute the initial Jacobian. However, a second model where the SPM has been shifted by \change{0.5\% of the pupil diameter} in the y-direction is used within the EFC controller to compute the electric field and apply the DM commands. In physical units, this shift is about 85microns, which is about 0.25 actuators when projected onto the DM pupil. Because of this unaccounted for error, $\beta$ scheduling and recomputation of the Jacobian become significant within the EFC controller to obtain adequate convergence. How often to switch regularization along with the values to switch between will depend on the coronagraph and the amount of model error. \change{Here, using a strict $\beta$ value of -2.5 and a relaxed $\beta$ value of -5 while alternating every 3 iterations} was found to be an effective schedule after trial and error. Additionally, a new Jacobian is recomputed with the nonlinear coronagraph model every 15 iterations because as DM commands accumulate, the \change{previous Jacobian becomes stale and EFC would diverge regardless of the $\beta$ schedule. After every relinearization, the $\beta$ schedule is restarted at the strict value of -2.5. Relinearizing more frequently at every 12 iterations was also simulated, but did not yield significant improvements in contrast or convergence.} 

While this use of $\beta$ scheduling and recomputation of the Jacobian can mitigate the effect of model errors for EFC, the controller requires more iterations to converge and becomes more complicated with $\beta$ values needing to be tuned according to the coronagraph and model. \change{The multiple relinearizations also increases the computational requirements of the controller, which can become an issue for a HWO since computations will be done on-board instead of using a ground-in-the-loop scheme\cite{pogorelyuk_computational_2022}}. Additional methods such as the expectation-maximization technique detailed in Sun et al.\cite{sun-system-id-2018} have been proposed to recover the state of a system and improve the model Jacobian, but this adds additional complexity to the controller that is not implemented for comparison here. 

\section{Monochromatic iEFC Results}
\label{sec:mono-iefc}
\begin{figure}[h]
    \centering
    \includegraphics[scale=0.6]{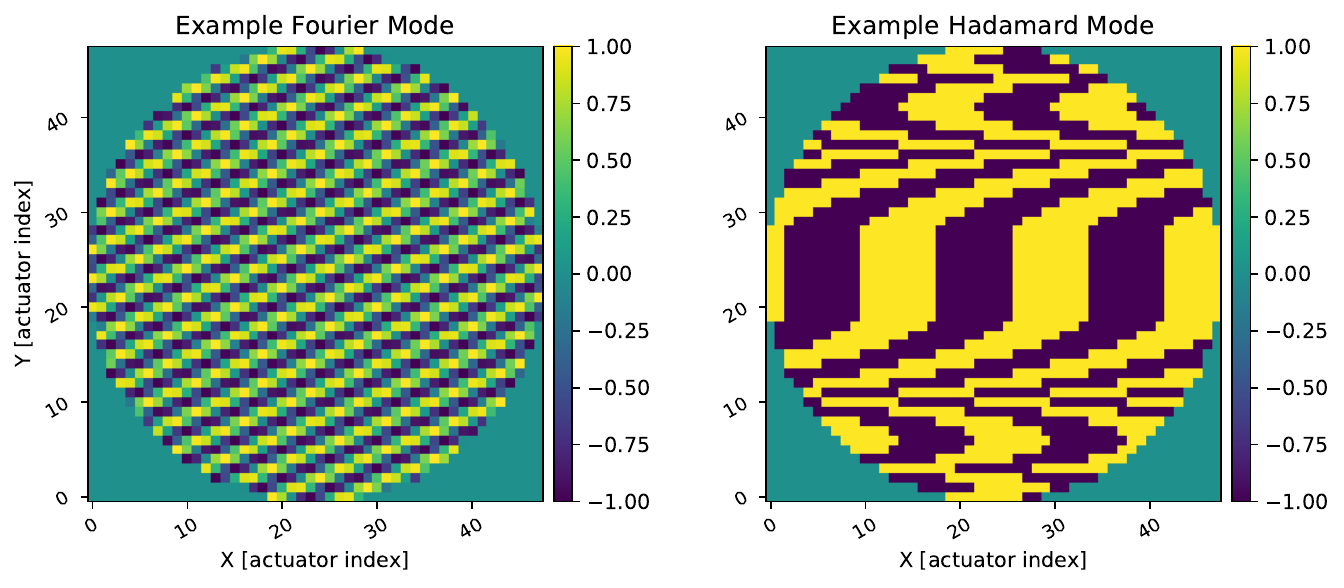}
    \caption{The images above depict examples of a Fourier mode and a Hadamard mode. The Fourier mode is a particular cosine corresponding to a specific spatial frequency that will be used within the control loop. The Hadamard modes are binary commands that span the entire vector space of DM commands.}
    \label{fig:example-modes}
\end{figure}

Using the SPC-WFOV mode at 825nm, the iEFC Jacobian is generated for three types of modes: Fourier modes, individual actuators, and Hadamard modes. Figure \ref{fig:example-modes} depicts examples of both a Fourier mode and a Hadamard mode. The Fourier modes are determined by discretizing the focal plane with a chosen spatial frequency sampling and generating cosine and sine commands corresponding to each of the desired spatial frequencies. The Hadamard modes on the other hand are generated based on the total number of actuators on a DM. 

\begin{figure}[h]
    \centering
    \includegraphics[scale=0.5]{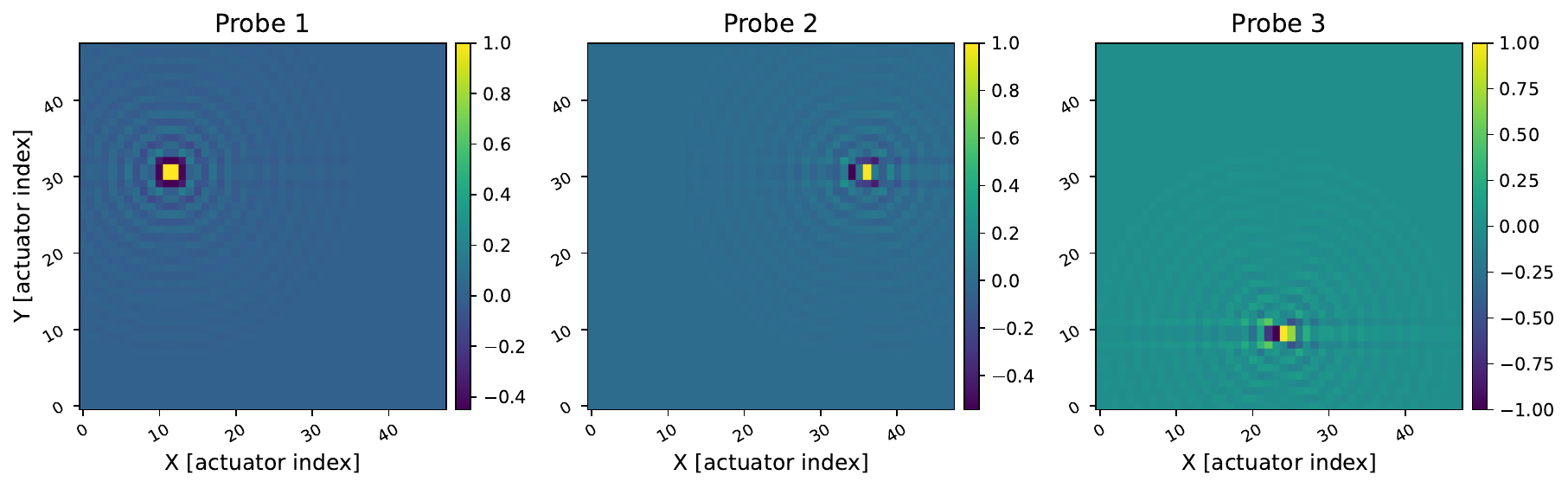}
    \caption{These probe commands are each a weighted sum of cosine and sine Fourier modes that have been shifted to a region of actuators not attenuated by the SPM. The first probe is solely a superposition of cosine Fourier modes (left), the second is an equal superposition of cosine and sine modes (middle), while the final probe is only a superposition of sine modes (right). The difference images of each probe are \change{measured to generate a response or compute modal coefficients.}}
    \label{fig:example-probes}
\end{figure}

For each modal basis, the same probes are used to generate the difference images of each calibration mode. These are analogous to the same probes that could be used for PWP. Initially, single actuator pokes are used for the probes as these were also used in Haffert et al\cite{haffert-iefc}. However, switching to the Fourier probes depicted in Figure \ref{fig:example-probes} improved the simulations with noise in Section \ref{sec:noisy-iefc} because the flux in the control region is increased, allowing for higher SNR and reduced calibration times. For consistency, these monochromatic simulations also implement the same Fourier probes where each DM command is a weighted sum of cosine and sine modes that span the desired control region. This helps to generate a relatively uniform response in the dark hole. The total sum for each probe is then shifted to a region of the DM where the actuators are not obstructed by the SPM. 

Overall, four sets of calibration modes were tested; two sets of Fourier modes, one set of single actuator pokes, and one set of Hadamard modes. For each set, the calibration amplitude ($a_{mode}$) was set to 5nm and the probe amplitude ($a_{probe}$) was set to 20nm. These values are chosen based upon the typical range of values used for testbed experiments with iEFC. After each calibration, iEFC is performed for 30 iterations because that is the desired value for the Roman Coronagraph HOWFSC plan\cite{zhou-cgi-howfsc-2020}.

\begin{figure}[h]
    \centering
    \includegraphics[scale=0.6]{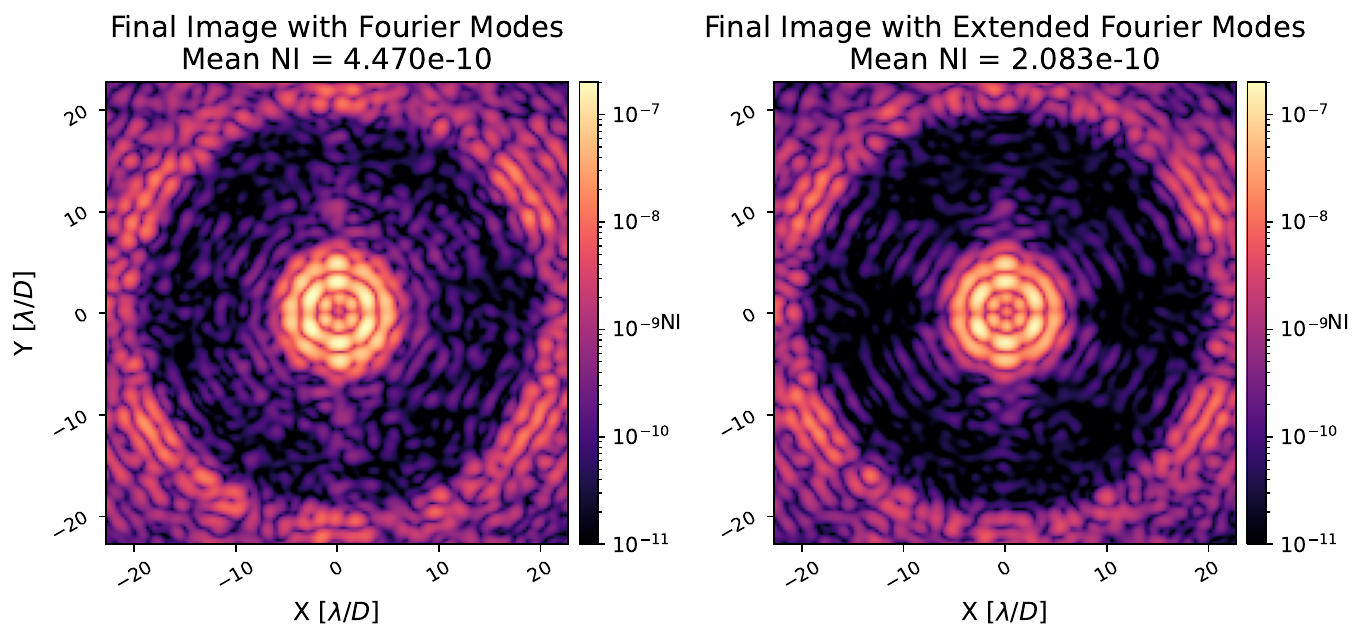}
    \caption{The iEFC solutions with Fourier modes using only the spatial frequencies of the control region only converge to a final contrast of $4.47\times10^{-10}$ while the solutions with extended Fourier modes reach $2.08\times10^{-10}$. This demonstrates that control of higher and lower spatial frequencies is essential for the HOWFSC method to perform optimally.}
    \label{fig:iefc-fourier}
\end{figure}

When it came to the Fourier modes, two sets were used because the initial set only considered spatial frequencies within the desired control region. This meant that the only controllable frequencies spanned from $5.4\lambda/D$ to $20.6\lambda/D$ with a sampling of $1\lambda/D$. The result after iEFC yielded a contrast of $4.47\times10^{-10}$. However, by using an expanded set of Fourier modes that spanned from $1\lambda/D$ to $23\lambda/D$, the final contrast reached is $2.08\times10^{-10}$. Figure \ref{fig:iefc-fourier} depicts the two results using both the constrained and extended Fourier modes. In total, the constrained set contained 2552 modes and the extended set contained 3480 modes.  

\begin{figure}[h]
    \centering
    \includegraphics[scale=0.6]{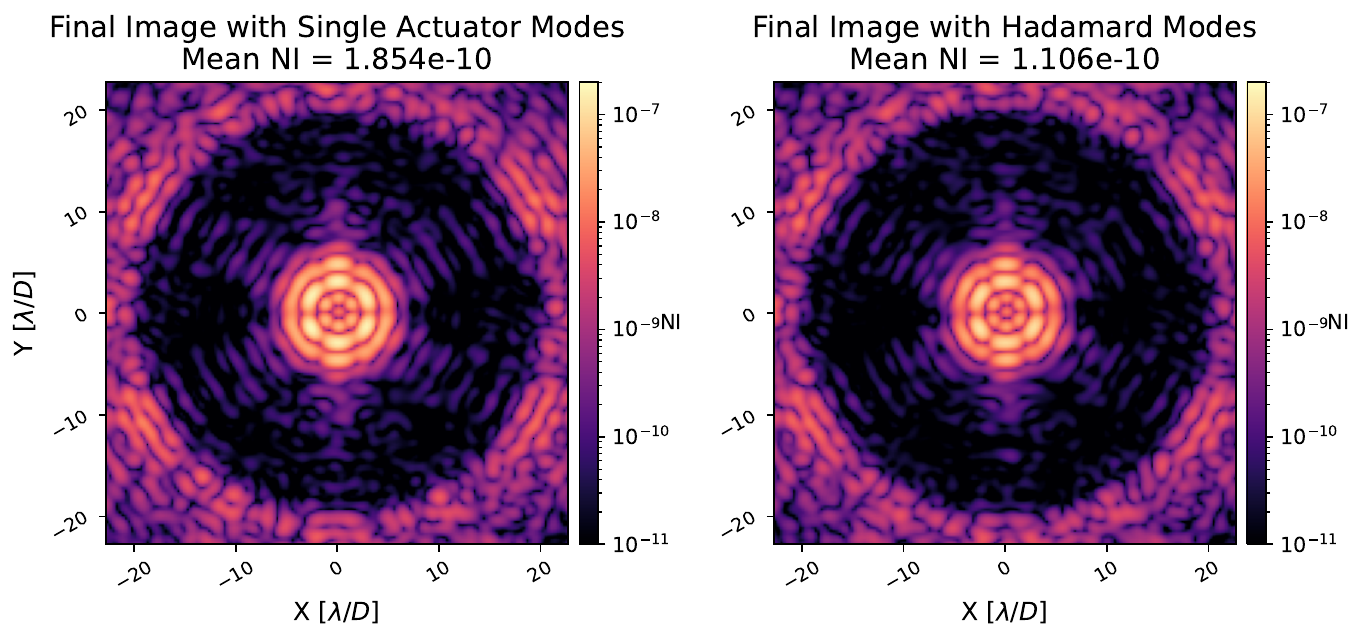}
    \caption{Using individual actuator modes, iEFC converged to a contrast of $1.85\times10^{-10}$. With Hadamard modes, the final contrast was slightly improved to a value of $1.11\times10^{-10}$.}
    \label{fig:iefc-poke-had}
\end{figure}

For the single actuator modes, the total number of actuators is culled to only the ones within the extent of the geometric pupil, which is about 46.3 actuators across the \change{diameter and totals 1804 actuators on each DM, or 3608 total modes. Using this basis results in a slight improvement over the extended Fourier modes as the final contrast reached is $1.85\times10^{-10}$.} Lastly, the Hadamard modes are selected by generating a set for each DM. As Hadamard modes are generated in powers of 2 and a single DM is assumed \change{to use 1804 actuators, a single DM uses 2048 Hadamard modes. Calibrating both DMs meant using a total of 4096 modes and the final mean contrast was $1.1\times10^{-10}$. Depicted in Figure \ref{fig:iefc-poke-had} are the final images for both the single actuator modes and the Hadamard modes.} 

\begin{figure}[h]
    \centering
    \includegraphics[scale=0.6]{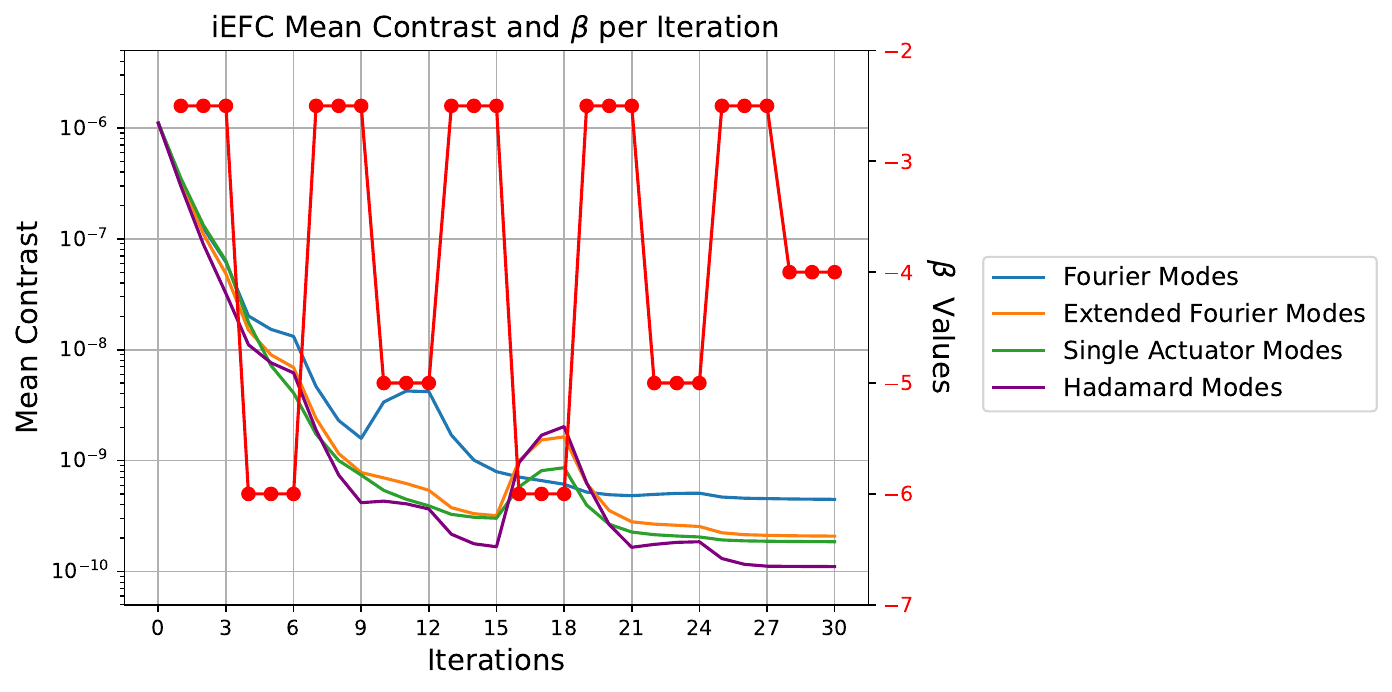}
    \caption{\change{Presented here are the mean contrasts and the respective $\beta$ values per iteration for each modal basis tested. While the margins between the various modal bases are narrow, it remains notable that the best contrast achieved was $1.11\times10^{-10}$ with Hadamard modes.}}
    \label{fig:mode-comps}
\end{figure}

For all simulations of iEFC, the control loop used the same $\beta$ schedule for one-to-one comparisons after tuning the values of the scheduler based upon what yielded adequate convergence within 30 iterations. Figure \ref{fig:mode-comps} directly compares the performance of iEFC using all calibration modes considered along with the $\beta$ schedule used. Similar to EFC, a strict value of -2.5 was found to be sufficient to converge without degrading contrast while the relaxed value oscillates from -6 to -5. The final $\beta$ used is -4 because that was the most relaxed value that could be used on the final iterations while converging and not requiring any more iterations with a value of -2.5. 

\begin{figure}[h]
    \centering
    \includegraphics[scale=0.525]{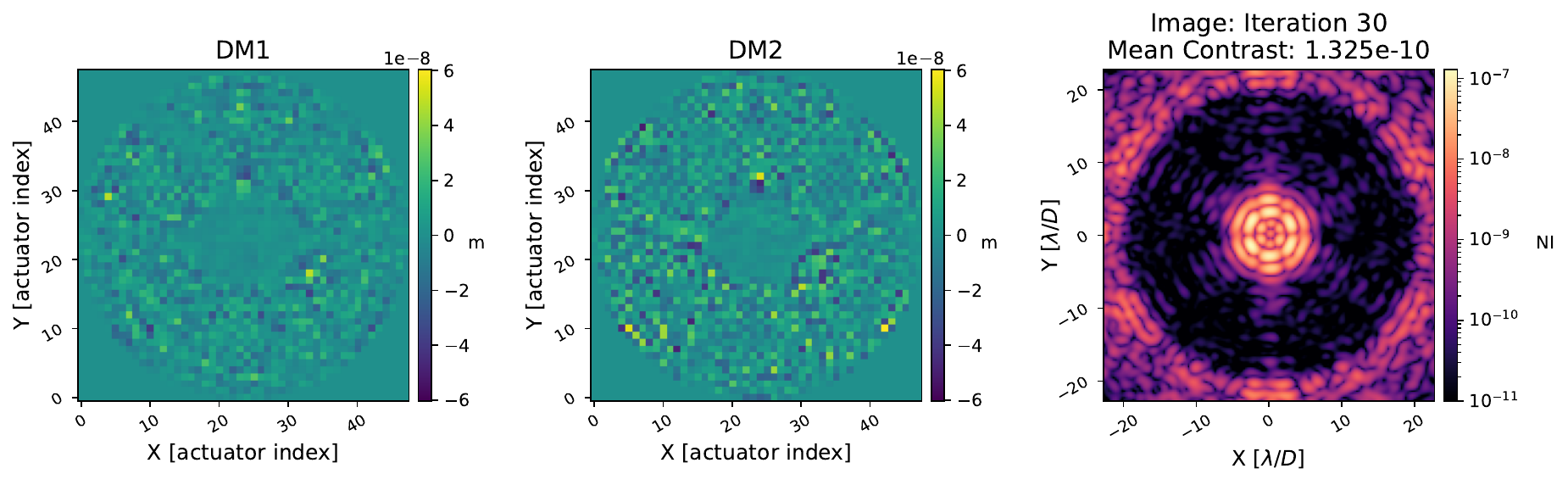}
    \caption{After simulating iEFC with the same SPM \change{shear} introduced in the EFC simulations, a 
    much smaller contrast degradation is noticed with the final contrast now being $1.33\times10^{-10}$.}
    \label{fig:iefc-perturbed}
\end{figure}

Although all results with iEFC attained similar convergence and final contrast values, the best result was with Hadamard modes. Further tuning of the $\beta$ values for each modal basis could improve the results individually, but for the one-to-one comparison done here, Hadamard modes proved the most effective. As such, this basis is chosen for the next simulation which includes the same lateral SPM shift that was introduced in the perturbed EFC simulation. The significant difference with iEFC is that the error is present during the calibration given all empirical data would inherently capture the perturbation. For this simulation, the same $\beta$ regularization schedule was used and the final \change{solutions depicted in Figure \ref{fig:iefc-perturbed}} achieved a slightly worse final contrast with a value of $1.33\times10^{-10}$. While the ideal scenario for EFC achieved the same contrast as the best case of iEFC within 15 iterations, this simulation demonstrates that iEFC is a preferable alternative in the scenario where the instrument is not well calibrated because EFC required 60 iterations with the increased computational complexity of relinearizing the Jacobian regularly. 

\section{Broadband iEFC Examples}
\label{sec:broad-iefc}
Given the Roman Coronagraph will have both narrowband filters chosen for wavefront control purposes and broader bandpasses for science observations, additional simulations have been performed for both a wavefront control filter and a science bandpass. Here, filter 4b is chosen as the wavefront control filter \change{with a central wavelength of 825nm and a FWHM of 3.6\%\cite{krists-bible}. For the science filter, a bandpass of 10\% is assumed.} 

To simulate a single broadband image, multiple wavefronts are propagated in the physical optics model with wavelengths spanning the chosen bandpass. The final image is an incoherent sum of the wavefronts at the image plane. For the narrowband simulations, 3 equally spaced wavelengths are selected to span the bandpass while 5 wavelengths are selected for the 10\% bandpass. Each wavefront is weighted equally in the final incoherent sum, so no variations in the filter transmittance are being considered. 

\begin{figure}[h]
    \centering
    \includegraphics[scale=0.48]{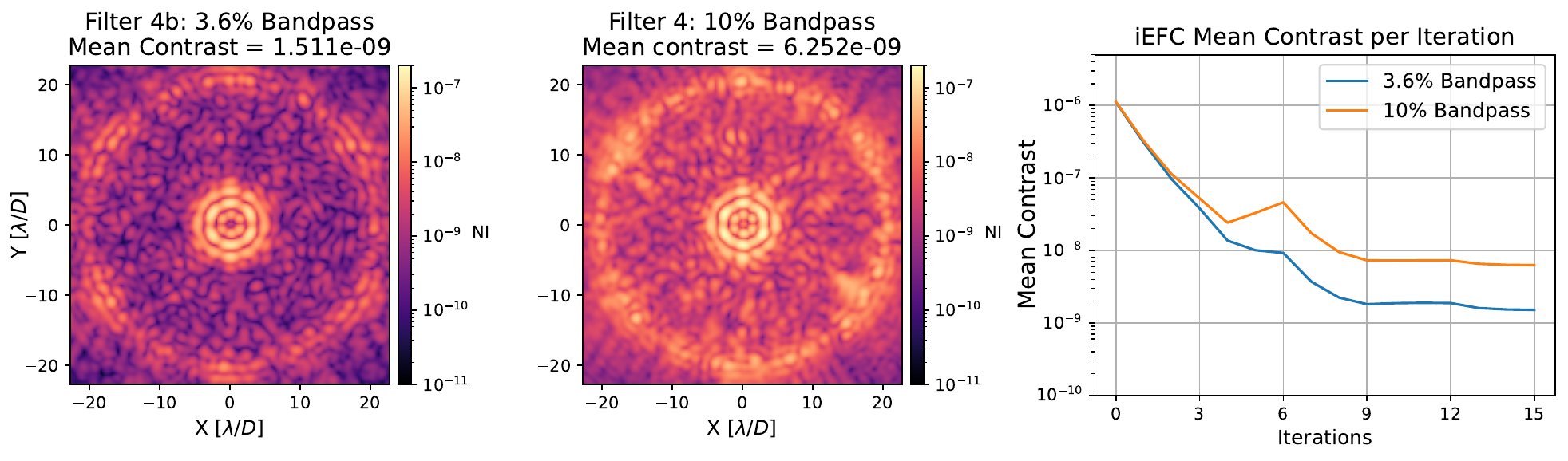}
    \caption{The final image using the single narrowband filter (left) illustrates that iEFC reached \change{$1.51\times10^{-9}$} contrast while the image using the 10\% bandpass (middle) has a contrast of $6.25\times10^{-9}$. As depicted in the plot of contrast per iteration (right), both converged in 15 iterations. More iterations were attempted while varying the regularization, but better results likely require additional \change{calibrations of iEFC.}}
    \label{fig:broadband figure}
\end{figure}

After calibrating iEFC for each bandpass, the control-loop is repeated with results depicted in Figure \ref{fig:broadband figure}. Given that Hadamard modes had the best performance in the monochromatic simulations, that was the chosen modal basis here. After iEFC is performed, the 3.6\% bandpass reaches a contrast of $1.51\times10^{-9}$ while the 10\% bandpass reaches a contrast of $6.25\times10^{-9}$. Importantly, these are the results after just 15 iterations because convergence was significantly slower after reaching these contrast values. 

This demonstrates that even in the full 10\% bandpass, iEFC can be capable of reaching below $10^{-8}$ contrast with the SPC-WFOV mode. Additional calibrations of iEFC after converging for a given bandpass could improve broadband iEFC results, but as explained in Section \ref{sec:noisy-iefc}, iEFC will likely require lengthy calibration times making recalibration \change{impractical}. 

\section{iEFC Simulations with Noise}
\label{sec:noisy-iefc}
Due to the empirical calibration of iEFC, noise will inherently be included in the Jacobian. To consider the impact of noise, these final simulations consider the same narrowband and broadband cases from Section \ref{sec:broad-iefc}, but flux estimates for a given reference star along with a model of Roman's EMCCD are used to generate more realistic images with photon and detector noise. 

Here, $\zeta$ Puppis is chosen as the reference star being utilized for wavefront control as it was in the Roman observing scenario simulations\cite{krist-spc-wfov-os11}. As described in Appendix \ref{app:blackbody}, the flux for $\zeta$ Puppis is computed at each wavelength that will be propagated for the chosen bandpass. \change{The amplitude of each wavefront is then weighted by the square root of the flux estimate for the respective wavelength and propagated to compute an image in units of photons/second/pixel. An additional factor of 0.5 is applied to the flux at the image plane to approximate the throughput losses from reflective surfaces throughout the optical train (note that this is based on a reflectivity of approximately 97.5\% for 25 reflective optics). Next, the image plane flux is} input into the emccd\_detect\footnote{\href{https://github.com/roman-corgi/emccd_detect}{https://github.com/roman-corgi/emccd\_detect}} software to simulate an image with EM gain and detector noise. The additional parameters chosen to model the EMCCD are provided in Table \ref{tab:emccd}.

\begin{table}[h]
    \centering
    \captionsetup{width=.8\textwidth}
    \caption{Below are the parameters used to model the EMCCD. The bias is chosen based on the current best estimate of read noise for the CGI EMCCD.}
    \begin{tabular}{|L{6cm}|L{3cm}|L{3cm}|} \hline 
    EMCCD Parameter & Quantity & Units \\ \hline
    Read Noise& 120 & \electron \\ \hline 
    Dark current & 8e-4 &  \electron/pixel/s \\ \hline
    Bias & 500  & \electron \\ \hline
    Full Well (Serial) & 100,000  & \electron \\ \hline
    Full Well (Image) &  60,000 & \electron \\ \hline
    Charge Induced Current (CIC) & 0.1 & \electron/pixel/frame \\ \hline
    Electrons per dn & 1 & \electron \\ \hline
    N Gain Registers & 604 & - \\ \hline
    Quantum Efficiency & 0.5  & - \\ \hline
    N-bits &  16 & - \\ \hline
    \end{tabular}
    \label{tab:emccd}
\end{table}

Again, only Hadamard modes are considered \change{to provide a direct comparison with the noiseless simulations. However, additional methods were implemented during the calibration to prevent both low SNR responses and responses with saturated pixels. The first of these methods is to implement a varying calibration amplitude for modes with a response that is distributed across more pixels in the image. Specifically, the response of each Hadamard mode is first estimated with the magnitude of the mode's Fourier transform. The maximum value of each estimated response is recorded such that the calibration amplitude can be scaled according to the ratio of the maximum for a given mode and the largest maximum recorded. Therefore, the mode with the largest maximum value in it's response uses a calibration amplitude of 5nm, but all others are multiplied by a scaling factor greater than 1.} 

\change{The second method used to calibrate both bandpasses is to average and stack frames with varying exposure times. This effectively increases the dynamic range of the images because pixels with large flux values do not get saturated at the lowest exposure time, but pixels with low flux values acquire a higher SNR when using the longer exposure times. For both the 3.6\% and 10\% bandpasses simulated, exposure times of 0.01s, 0.05s, and 0.1s are stacked to generate the final image. The number of frames for each respective exposure time are 10, 4, and 2. The total integration time for each calibration frame is then 0.5s, so to calibrate all 4096 Hadamard modes is estimated to require about 6.8hrs of data acquisition.} Note that to perform full broadband control using all three narrowband wavefront control filters will require $3\times$ the integration time, so a total of 20.4hrs, but this is not simulated here. \change{Here, the EM gain used for all frames in the 3.6\% bandpass is 250 and the EM gain for frames in the 10\% bandpass is 100.}

\begin{figure}[h]
    \centering
    \includegraphics[scale=0.5]{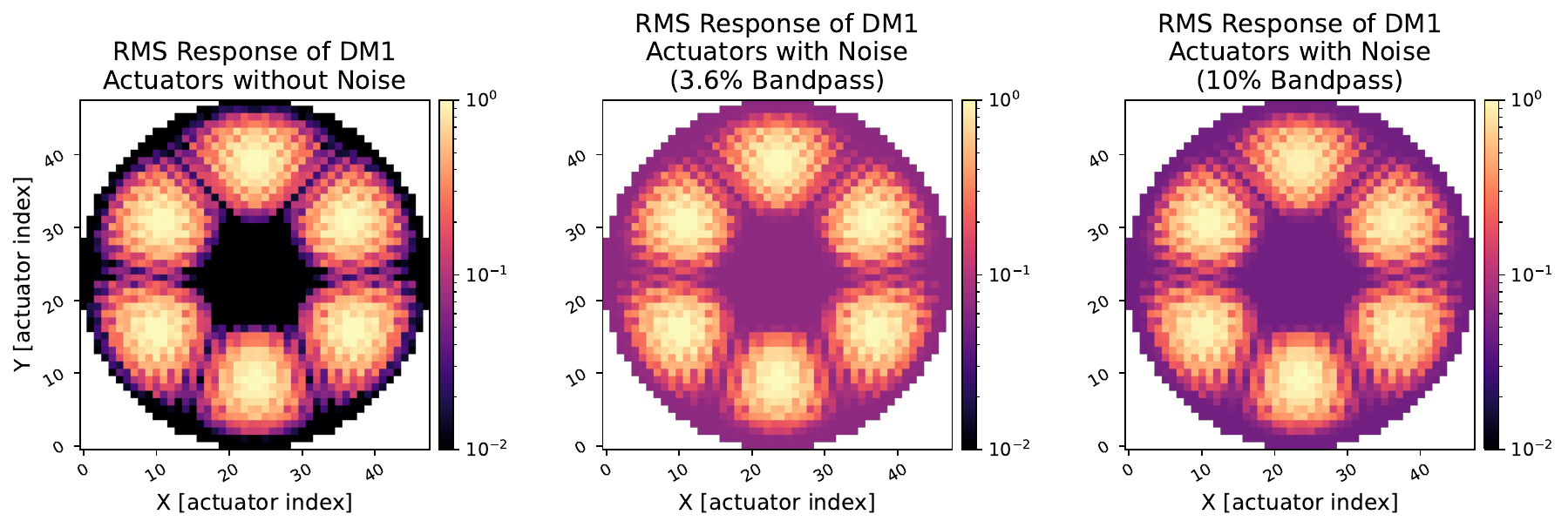}
    \caption{After calibrating iEFC while including noise, the RMS response of DM1 actuators provides a useful visualization of the quality of calibration. On the left is the RMS response with a noiseless calibration while the middle and right show the RMS responses after calibrating the 3.6\% and 10\% bandpasses with noise. The constant bias of the obstructed actuators is a direct result of the noise from calibration. Here, the simulation of the 10\% bandpass has a slightly lower bias than the 3.6\% bandpass because the additional photons improved the SNR.}
    \label{fig:rms-responses}
\end{figure}

Figure \ref{fig:rms-responses} depicts a useful visualization for the quality of the calibration. Each image in Figure \ref{fig:rms-responses} illustrates the normalized RMS response of each actuator on DM1. In the case without any noise, the obstructed actuators have a considerably attenuated response while the two scenarios with noise have a relatively constant bias that is a direct result from noise during calibration. When using the 10\% bandpass, the bias of the obstructed actuators is slightly less due to the higher SNR achieved by capturing more photons.

\begin{figure}[h]
    \centering
    \includegraphics[scale=0.48]{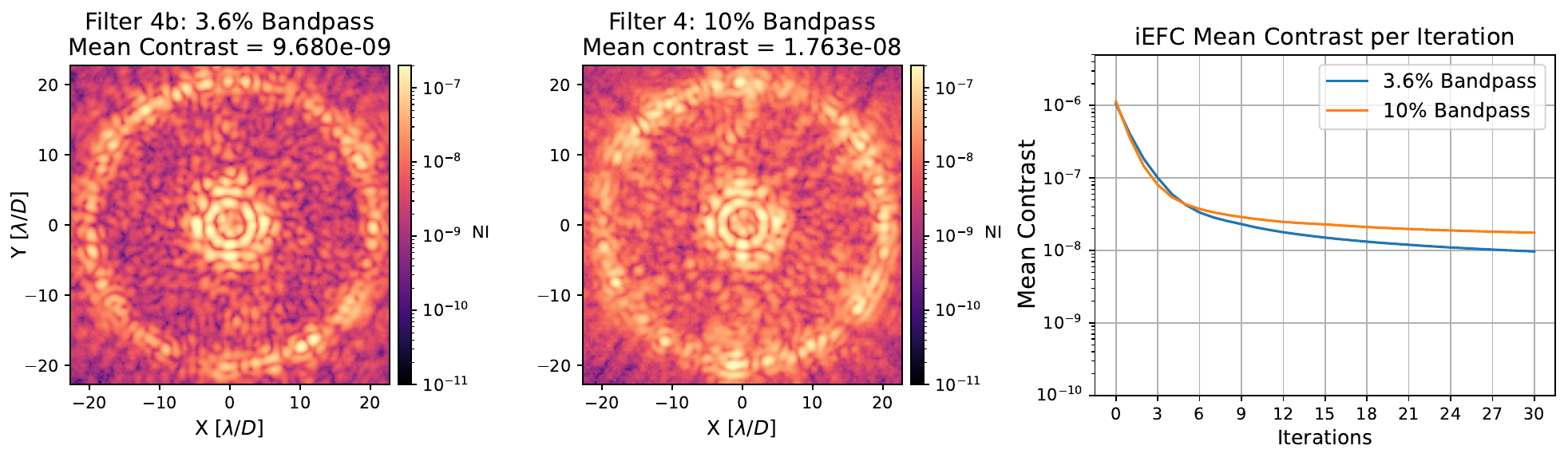}
    \caption{With noise implemented, iEFC reached \change{$9.68\times10^{-9}$ using the 3.6\% bandpass (left) while $1.76\times10^{-8}$ was reached with the 10\% bandpass (middle). Here, 30 iterations were used for each simulation because the noise from the calibration degraded the convergence for each bandpass.}}
    \label{fig:noisy-iefc}
\end{figure}

Because of the noise accumulated during calibration, we see additional degradation in the performance of iEFC as presented in Figure \ref{fig:noisy-iefc}. With the narrowband filter, the contrast converged to \change{$9.68\times10^{-9}$} while the 10\% bandpass reached \change{$1.76\times10^{-8}$}. Each simulation only used 30 iterations because the convergence of iEFC became stagnant even when the regularization value was varied. 

These simulations demonstrate the disadvantage of iEFC which is the noise accumulated during calibration. Possible mitigation strategies could be a smarter choice of EMCCD parameters including the use of higher gain values and photon counting to reduce detector noise, using a brighter reference star (but this will depend on the desired target star and stability of the optical assembly), using longer integration times for calibration, or choosing a more optimal set of modes for the controller. In any case, additional techniques should be investigated to mitigate the impact of noise in the calibration of iEFC if it is to be implemented by Roman.

\section{Conclusion}
\label{sec:conclusion}
The iEFC algorithm is an exciting method for the high-contrast imaging community. Like EFC, iEFC is naturally extended to coronagraphs with two DMs and capable of digging annular dark holes. This makes iEFC a potential method for the Roman Coronagraph and future coronagraphs for the HWO. Here, the efficacy and potential benefits of iEFC have been investigated specifically for the SPC-WFOV mode due to the large linear regime of the DMs. The simulations demonstrate that contrasts comparable to those of EFC can be achieved in the ideal monochromatic scenario and that Hadamard modes are the most effective for the control-loop. More significantly, iEFC would reduce the risk of unknown system perturbations that would otherwise require a more complicated $\beta$ scheduling method, more iterations, and greater computational complexity due to relinearizations of the Jacobian. By inherently including the state of the instrument during calibration, iEFC does not need to be relinearized for coronagraphs with large linear regimes such as the SPC-WFOV mode or a vortex coronagraph as simulations here demonstrate that adequate convergence can be obtained within 30 iterations. 

Additional simulations demonstrate that iEFC can be calibrated and performed with both a narrowband filter and a broadband filter, but the performance of iEFC degrades due to the incoherent \change{superposition of wavelengths. Further simulations including the EMCCD model predict that contrasts of $10^{-8}$ can be achieved within 30 iterations after calibrating for an estimated time of 6.8hrs when using $\zeta$ Puppis as the reference star.} Further investigations including more optimal EMCCD parameters, smarter choice of calibration modes, and improved calibration techniques could reduce the calibration time and improve the final contrast, but other factors such as system drift during calibration should also be considered in the future. Given the simulation results for the time being, the SPC-WFOV mode is a suitable instrument to perform iEFC tests during the latter stages of the Roman mission in order to inform the WFSC strategies for future coronagraphs such as those considered for the HWO.

\subsection*{Acknowledgments}
Portions of this research were supported by funding from the Technology Research Initiative Fund (TRIF) of the Arizona Board of Regents and by generous anonymous philanthropic donations to the Steward Observatory of the College of Science at the University of Arizona. Additional support was received through the WFIRST Science Investigation team prime award \#NNG16PJ24C. This study utilized High Performance Computing (HPC) resources supported by the University of Arizona TRIF, UITS, and Research, Innovation, and Impact (RII) and maintained by the UArizona Research Technologies department. This research made use of community-developed core Python packages, including: POPPY\cite{perrin-poppy-2017}, Astropy \cite{the_astropy_collaboration_astropy_2013}, Matplotlib \cite{hunter_matplotlib_2007}, SciPy \cite{jones_scipy_2001}, Ray\cite{moritz_ray_2018}, and the IPython Interactive Computing architecture \cite{perez_ipython_2007}. Lastly, A.J. Riggs, Jaren Ashcraft, Kevin Derby, Leonid Pogorelyuk, Aaron Goldtooth, and Ranger Maxwell have contributed through helpful discussions regarding simulations and wavefront control throughout the span of this research. 

\section{Code, Data, and Materials Availability}

All the code for simulations is made available through public repositories. The repository containing the end-to-end diffraction model constructed with POPPY can be found on Zenodo at \href{https://zenodo.org/record/8302380}{https://zenodo.org/record/8302380}\cite{milani-cgi-phasec-poppy}. Similarly, the repository containing the simulation material can be found at \href{https://zenodo.org/records/8302359}{https://zenodo.org/records/8302359}\cite{milani-roman-cgi-iefc}. 

The associated data that is required for the diffraction model along with the data obtained using iEFC can be found at \href{https://github.com/kian1377/cgi_phasec_poppy_data}{https://github.com/kian1377/cgi\_phasec\_poppy\_data}. 

\appendix
\section{Definitions}
\label{app:defs}
The variables used for the formalism of EFC and iEFC are summarized here. The shapes of the matrices and vectors are also included to clarify how the algorithms are implemented (Table \ref{tab:vars}).

\begin{table}[H]
    \centering
    \caption{This table contains the significant variables used for the descriptions of EFC and iEFC. \change{Note that the Jacobian array sizes are assumed to be for a single wavelength or bandpass.}}
    \begin{tabular}{|L{1.75cm}|L{6.25cm}|L{7cm}|} \hline
        Variable & Description & Type and Shape \\ \hline
        $A(x,y)$ & Transverse amplitude aberrations of the pupil plane scalar electric field & Scalar \change{field} \\ \hline
        $\Phi(x,y)$ & Transverse phase aberrations of the pupil plane scalar electric field & Scalar \change{field} \\ \hline
        $\Phi_{DM} (x,y)$ & Transverse phase applied by the DM surface & Scalar \change{field} \\ \hline
        $N_{pix} $ & Number of pixels within the desired control region of the focal plane & Scalar \\ \hline
        $N_{modes} $ & Number of modes forming the basis of DM commands & Scalar \\ \hline
        $N_{probes} $ & Number of probe commands used for PWP or iEFC & Scalar\\ \hline
        $M_{modes} $ & Matrix containing the chosen modal basis  & Matrix with shape $[N_{act}^2\times N_{modes}]$ for 1 DM or $[2 N_{act}^2\times N_{modes}]$ for 2 DMs\\ \hline
        $\mathbf{E_{im}}$ & Total electric field of each pixel within the control region & Vector with length $2N_{pix}$ \\ \hline
        $\mathbf{E_{ab}}$ & Aberrated electric field within the control region sensed for EFC & Vector with length $2N_{pix}$ \\ \hline
        $\boldsymbol{\delta}$ & \change{Difference images for a set of probe commands} & Vector with length $N_{probes} N_{pix}$ \\ \hline
        $G_{EFC} $ & Jacobian for EFC & Matrix with shape $[2N_{pix}\times N_{modes}]$\\ \hline
        $G_{IEFC} $ & Jacobian for iEFC & Matrix with shape $[N_{probes}N_{pix}\times N_{modes}]$\\ \hline
        $\mathbf{A} $ & Individual actuator \change{commands} & Vector with length $N_{acts}$\\ \hline
        $\mathbf{m_c}$ & Modal coefficients for the chosen modal basis for DM commands & Vector with length $N_{modes}$ \\ \hline
        $M_{control}$ & The control matrix \change{computed/measured} for the respective control-loop & Matrix with shape $[N_{modes}\times 2N_{pix}]$ for EFC or $[N_{modes}\times N_{probes}N_{pix}]$ for iEFC\\ \hline
    \end{tabular}
    \label{tab:vars}
\end{table}

\section{Blackbody Flux Calculations}
\label{app:blackbody}

\begin{figure}[h]
    \centering
    \includegraphics[scale=0.6]{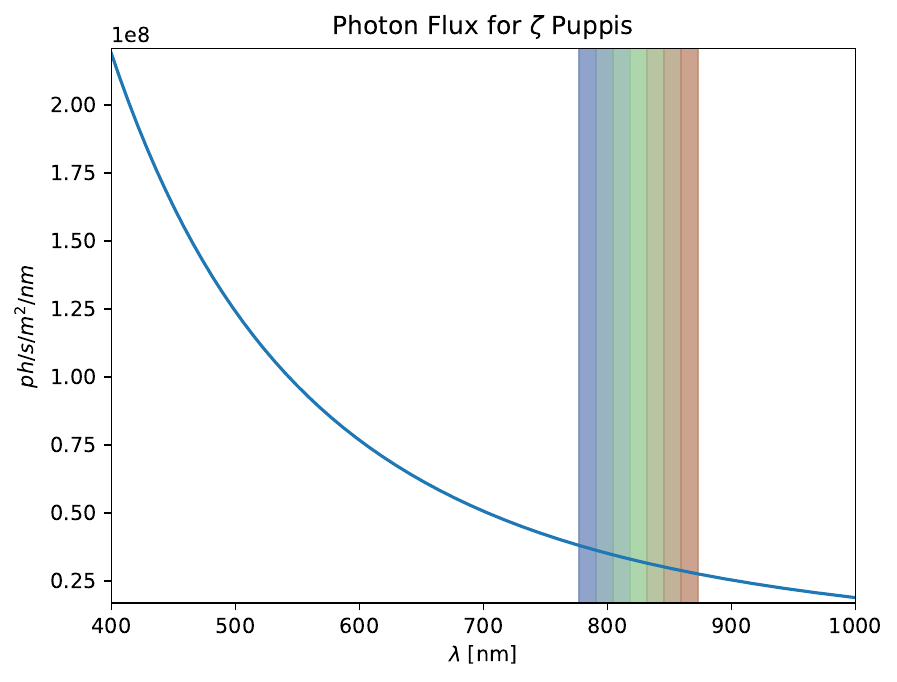}
    \caption{This is an example of the spectral flux computed for $\zeta$ Puppis using blackbody equation. The highlighted regions illustrate the sub-bandpasses integrated to estimate flux-values for individual wavelengths centered at each sub-bandpass. Note that the peak of the blackbody is located at about 200nm due to the temperature of the star.}
    \label{fig:blackbody}
\end{figure}

To include estimates of flux, the blackbody equation 
\begin{equation}
\label{eq:blackbody}B_{\lambda} = \frac{2hc^2}{\lambda^5} \frac{1}{\exp{[\frac{hc}{\lambda k_B T}]} - 1}.\end{equation}
\noindent is used to calculate the spectral radiance for a chosen star in units of W/m$^2$/sr/nm\cite{grant-field-guide-radiometry-2011}. Here, $\zeta$ Puppis is chosen as the reference star because it is a bright star planned to be used for HOWFSC in the Roman Observing Scenarios\cite{krist-spc-wfov-os11}. The spectral radiance of $\zeta$ Puppis is calculated using the temperature of $40,000K$\cite{zeta-pup-temp}. The spectral irradiance is then calculated using the solid angle of the star calculated with the equation \begin{equation} 
\Omega=2\pi\left(1-\frac{\sqrt{d^2-R^2}}{d}\right)\text{.}
\end{equation} 
\noindent Here, $R$ is the radius of the star and $d$ is the distance to the star, both of which are included in Table \ref{tab:zeta-pup-params}. The spectral photon flux is calculated by converting the spectral irradiance using $E=hc/\lambda$. Lastly, the spectral flux is integrated over several sub-bandpasses to estimate the flux for each wavelength that is propagated. \change{Figure \ref{fig:blackbody} presents an example of the spectral photon flux computed for $\zeta$ Puppis and 7 sub-bandpasses that would be integrated to estimate the photon flux for those 7 wavelengths.} 

\begin{table}[h]
    \centering
    \caption{The parameters used to calculate the spectral radiance of $\zeta$ Puppis are presented below along with the final solid angle used to calculate flux.}
    \begin{tabular}{|c|c||} \hline
       Temperature ($T$) & 40,000K\cite{zeta-pup-temp} \\ \hline 
       Radius ($R$) & 14$R_\odot$\cite{zeta-pup-radius} \\ \hline 
       Distance ($d$) & 300parsec\cite{zeta-pup-radius} \\ \hline 
       Solid Angle ($\Omega$) & $1.362\times10^{-17}$sr \\ \hline
    \end{tabular}
    \label{tab:zeta-pup-params}
\end{table}

\bibliography{report}   
\bibliographystyle{spiejour}   

\vspace{2ex}\noindent\textbf{Kian Milani} received his BS in Optical Engineering from the Wyant College of Optical Sciences in 2020. Now a PhD candidate in the same college, his work focuses on physical optics simulations and wavefront control techniques for coronagraphic instruments. 

\vspace{1ex}
\noindent Biographies and photographs of the other authors are not available.


\end{spacing}
\end{document}